\documentclass[conference]{IEEEtran}
\IEEEoverridecommandlockouts
\usepackage{setspace}
\usepackage{cite}
\usepackage{amsmath,amssymb,amsfonts}
\usepackage{multirow}
\usepackage[linesnumbered,ruled]{algorithm2e}
\usepackage{array}
\usepackage{url}
\usepackage{verbatim}
\usepackage{stfloats}
\usepackage{graphicx}
\usepackage{dirtree}
\usepackage{subfigure}
\usepackage{float}
\usepackage{textcomp}
\usepackage{xcolor}
\usepackage{soul}
\usepackage{booktabs}
\usepackage{bm}
\def\BibTeX{{\rm B\kern-.05em{\sc i\kern-.025em b}\kern-.08em
		T\kern-.1667em\lower.7ex\hbox{E}\kern-.125emX}}
\usepackage{balance}
\definecolor{matlabyellow}{rgb}{0.9290,0.6940,0.1250}
\begin{document}
	\sethlcolor{yellow} 
	
	\title{OpenPathNet: An Open-Source RF Multipath Data Generator for AI-Driven Wireless Systems}

	\author{
		\IEEEauthorblockN{Lizhou Liu, Xiaohui Chen, Wenyi Zhang$^{*}$}

		\IEEEauthorblockA{Department of Electronic Engineering and Information Science,\\
		University of Science and Technology of China, Hefei, Anhui 230027, China\\
		Email: liulizhou@mail.ustc.edu.cn; cxh@ustc.edu.cn; wenyizha@ustc.edu.cn}

		\thanks{This work was supported in part by the National Natural Science Foundation of China through Grant 62231022. Corresponding author: Wenyi Zhang.}
	}
	
	\maketitle

	\begin{abstract}
		The convergence of artificial intelligence (AI) and sixth-generation (6G) wireless technologies is driving an urgent need for large-scale, high-fidelity, and reproducible radio frequency (RF) datasets. Existing resources, such as CKMImageNet, primarily provide preprocessed and image-based channel representations, which conceal the fine-grained physical characteristics of signal propagation that are essential for effective AI modeling. To bridge this gap, we present OpenPathNet, an open-source RF multipath data generator accompanied by a publicly released dataset for AI-driven wireless research. Distinct from prior datasets, OpenPathNet offers disaggregated and physically consistent multipath parameters, including per-path gain, time of arrival (ToA), and spatial angles, derived from high-precision ray tracing simulations constructed on real-world environment maps. 
		By adopting a modular, parameterized pipeline, OpenPathNet enables reproducible generation of multipath data and can be readily extended to new environments and configurations, improving scalability and transparency. 
		The released generator and accompanying dataset provide an extensible testbed that holds promise for advancing studies on channel modeling, beam prediction, environment-aware communication, and integrated sensing in AI-enabled 6G systems. 
		The source code and dataset are publicly available at \url{https://github.com/liu-lz/OpenPathNet}.
	\end{abstract}
	

	\section{Introduction}

	In sixth-generation (6G) mobile communication networks, dense infrastructure deployment, large antenna arrays, wide bandwidths, and intelligent network architectures are anticipated to be defining technological features \cite{6G_1}. Accurately characterizing the spatial distribution and propagation behavior of radio frequency (RF) signals is crucial for extracting and utilizing channel state information (CSI). High-resolution RF maps enable efficient analysis and prediction of signal distributions, thereby playing a key role in enhancing the quality of service (QoS) of wireless communication systems.

	However, constructing a complete and accurate RF map remains a challenging task. Traditional empirical propagation models, such as the Okumura-Hata model \cite{Hata}, struggle to capture complex propagation behaviors in urban environments. Ray tracing methods \cite{Sionna} achieve higher accuracy but require substantial computational resources, while interpolation-based approaches (e.g., Kriging \cite{Kriging}) are efficient yet prone to significant errors under sparse or non-uniform sampling conditions.

	With the rapid advancement of artificial intelligence (AI), deep learning-based approaches for constructing RF maps have attracted significant research interest. Recent methods, such as RadioUNet \cite{UNet} and diffusion-based frameworks \cite{diffusion}, have shown progress in predicting signal strength or path loss. Nevertheless, these methods typically focus on a single physical quantity and overlook the multipath effects that dominate wireless propagation. Fundamentally, multipath propagation determines signal fading, interference behavior, and spatial correlation. Several studies have explored this direction, such as~\cite{multipath1}, which constructed a multipath RF map, and~\cite{multipath2}, which leveraged multipath information for high-precision localization.

	Despite these efforts, there remains a scarcity of high-fidelity datasets capable of systematically representing multipath characteristics. Existing public datasets \cite{DeepMIMO} -\cite{DeepSense} exhibit two primary limitations. First, they typically provide only aggregated channel responses or power distributions, often presented as image-based representations rather than the explicit path-level multipath parameters essential for multiscale modeling and physical interpretability. Second, most datasets are released as static collections rather than parameterized generation frameworks, thereby constraining reproducibility and scalability.

	To address these limitations, we present OpenPathNet as a configurable toolchain that, based on real-world environments, constructs simulation scenes and generates structured path-level multipath parameter sets through a unified and fully open workflow. Lightweight configuration files further enable efficient data generation for new environments and varied deployment settings (e.g., transceiver placement, carrier frequency, and antenna parameters). Rather than releasing task-specific representations, the pipeline outputs explicit path-level channel parameters, allowing researchers to construct customized inputs for different learning targets while retaining physical interpretability. Alongside the toolchain, we publicly release a reference dataset generated from representative real-world scenes using the proposed generator. Each sample provides explicit per-path physical parameters, including path gain, time of arrival (ToA), and spatial angles. This combination of a released dataset and a configurable generator facilitates scalable multipath data construction and transparent reuse by the community.

	\section{Related RF Datasets}

	With the rapid integration of AI into wireless communications, the demand for high-fidelity, large-scale RF datasets that accurately capture physical propagation characteristics has grown substantially. Recently, several RF datasets tailored for AI-driven research have emerged, playing a crucial role in advancing intelligent communication, sensing, and modeling. This section reviews and analyzes representative RF datasets, examining their design concepts, strengths, and limitations, and subsequently outlining the motivation for developing OpenPathNet.

	\subsection{DeepMIMO}

	DeepMIMO \cite{DeepMIMO} is a highly flexible and parameterized dataset designed for deep learning research in wireless communications. It employs ray tracing simulations to generate realistic channel data, allowing researchers to configure a wide range of parameters, including antenna settings, system bandwidth, number of paths, carrier frequency, and user distribution. This flexibility makes DeepMIMO a versatile tool for various tasks such as beam prediction, channel estimation, and localization. Notably, DeepMIMO provides rich RF information, including detailed multipath components with their corresponding delay, gain, and angular characteristics for each propagation path. However, a primary limitation is its reliance on a fixed set of predefined scenarios. Users are restricted to adjusting parameters within these provided environments, lacking the ability to import or construct customized scenes tailored to specific research needs..

	\subsection{RadioMapSeer}

	RadioMapSeer \cite{radiomapseer} is a large-scale radio map dataset designed for dense urban environments. It provides high-quality RF propagation samples for deep learning research. The dataset is built on ray tracing simulations that generate pathloss and ToA data across multiple city layouts and frequency bands. 
	By adopting a vision-driven organization strategy, RadioMapSeer converts pathloss values into two-dimensional (2D) images, thereby lowering the modeling barrier and facilitating the use of computer vision models for propagation prediction.
	It also facilitates vision-based wireless learning. However, this design introduces limitations. 
	Specifically, pathloss values are truncated and linearly scaled to 8-bit grayscale, a quantization process that restricts the dataset's utility for precise multidimensional modeling.
	When propagation characteristics are rasterized into pathloss maps, detailed multipath parameters, such as angular information and non-dominant path components, are lost. This hinders the analysis of temporal and spatial structures. Finally, RadioMapSeer is a largely closed framework. As a static dataset, it does not support new environment customization or reproducible ray tracing configurations. 
	Consequently, while it is well suited for macroscopic propagation modeling, it is less applicable to AI models requiring fine-grained, signal-level information.

	\subsection{CKMImageNet}
	CKMImageNet \cite{CKM} reformulates channel prediction into a vision-based learning task by representing wireless propagation environments as grayscale images. The approach enables wireless researchers to leverage deep neural architectures and training strategies from computer vision, thereby enhancing AI-driven channel characterization and beam prediction. 
	In terms of data diversity, CKMImageNet covers a broad range of propagation scenarios, utilizing high-precision ray tracing to generate structured training and testing datasets.
	However, its reliance on image-based representations rather than explicit multipath parameters presents a significant limitation. The process of mapping high-dimensional multipath data into 2D images inevitably aggregates and quantizes physical details, constraining in-depth propagation analysis. Furthermore, CKMImageNet is also not fully open source. The absence of complete modeling scripts and generation pipelines restricts researchers from customizing environments or adjusting parameters. Consequently, while CKMImageNet serves as a valuable benchmark for vision-based channel learning, it is less suitable for studies requiring high reproducibility, scalability, and physical fidelity.

	\subsection{RadioDiff-3D}

	RadioDiff-3D \cite{3D} is a large-scale 3D radio map dataset providing pathloss, angle of departure (AoD), and ToA information via image-based representations. Unlike existing 2D datasets, it enables fully 3D channel modeling and supports altitude-aware wireless applications. However, RadioDiff-3D is generated under fixed simulation scenarios and parameter settings, which prevents flexible extension to user-defined environments or custom propagation configurations. Furthermore, access to the underlying physical channel parameters remains unavailable.

	\subsection{Sensiverse}
	Sensiverse \cite{Sensiverse} is a large-scale dataset dedicated to integrated sensing and communication (ISAC), offering RF channel samples, 3D environmental models, and scripts for multi-scenario simulations. However, its channel data is provided as channel frequency responses rather than explicit per-path multipath parameters. Additionally, it is released as a static dataset, limiting the flexibility for scenario customization.

	\subsection{DeepSense 6G}
	DeepSense 6G \cite{DeepSense} is a real-measurement-based multimodal sensing and communication dataset covering diverse real-world environments. Nevertheless, it provides only received signal power on the communication side, lacking detailed multipath characteristics. Moreover, its reliance on complex data acquisition with specialized hardware platforms constrains both reproducibility and scalability.

	Based on the above analysis, existing RF datasets have played important roles in specific scenarios and tasks. However, they exhibit common intrinsic limitations that hinder broader support for AI-driven wireless communication research. One fundamental issue lies in the granularity of data representation. Most datasets provide highly aggregated or post-processed channel information rather than explicit path-level multipath parameters. Such representations can obscure key physical details of propagation, making it difficult for AI models to directly relate environment geometry to discrete multipath components. Another critical limitation is the lack of flexibility regarding the data-generation process. Most existing datasets are released as static collections of pre-generated results, rather than open-source and parameterized generation frameworks. As a result, researchers are often unable to incorporate custom environments, adjust simulation settings, or systematically extend the data to new environments and configurations, which constrains reproducibility, scalability, and adaptability. To address these gaps in data representation and framework openness, OpenPathNet is presented.

	\section{Dataset Design and Generation Framework} \label{section3}

	The core objective of OpenPathNet is to create a reproducible, scalable, and physically consistent framework for generating explicit path-level multipath data for AI-driven wireless communication research. Unlike conventional datasets that rely on closed simulators or empirical channel models, OpenPathNet focuses on end-to-end interpretability and physical-model consistency. This focus allows researchers to systematically evaluate model generalization, adherence to physical principles, and robustness on a unified foundation. This section elaborates on the framework's core modules and the automated pipeline enabling large-scale data generation.

	\subsection{Framework Overview and Design Principles} \label{section3-1}
	OpenPathNet is built on three design pillars: geographical and parametric configurability, physical consistency, and reproducibility. It generates per-path radio frequency data, not just aggregated channel responses or image-based representations. Unlike datasets that only provide channel gain or path loss, OpenPathNet preserves the fidelity and completeness of path-level observables such as complex gain, ToA, and multidimensional angular features. By minimizing reliance on empirical assumptions, such as stationarity or Gaussian approximations, the framework offers physically grounded data for learning-based channel modeling, wireless sensing, and multimodal environmental understanding.

	Inspired by Geo2SigMap \cite{geo2sigmap}, we developed an automated toolchain that enables flexible and scalable scene construction. Leveraging this toolchain, OpenPathNet further adopts a modular architecture integrating three essential components: environment modeling, ray tracing simulation, and data processing.
	The environment modeling module supports both automated import of city-scale 3D environments from geographic information systems and procedural synthesis of structured geometries, addressing diverse spatial coverage needs. Using automated scripts that combine \textit{OpenStreetMap}\footnote{https://www.openstreetmap.org/} and \textit{Blender}\footnote{https://www.blender.org/}, users can seamlessly generate simulation-ready 3D scenes from real-world latitude and longitude coordinates. When geographical data are unavailable, the framework enables procedural block generation through a parametric geometry engine, ensuring spatial continuity and statistical diversity for large-scale synthesis.

\begin{figure}[htbp]
		\centering
		\subfigure[] {\label{a}\centering\includegraphics[width=0.29\columnwidth]{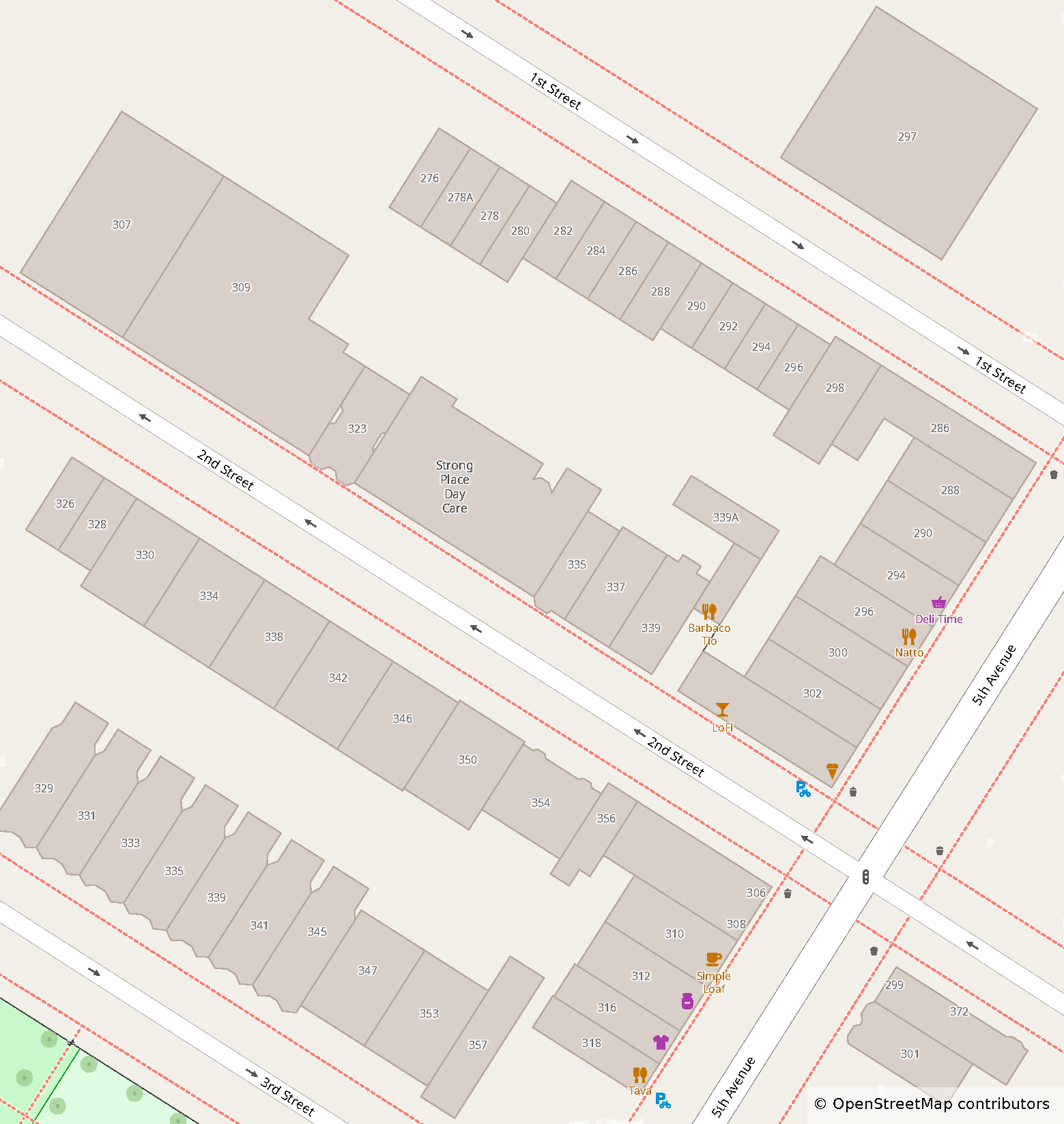}%
		}
		\subfigure[] {\label{b}\centering\includegraphics[width=0.35\columnwidth]{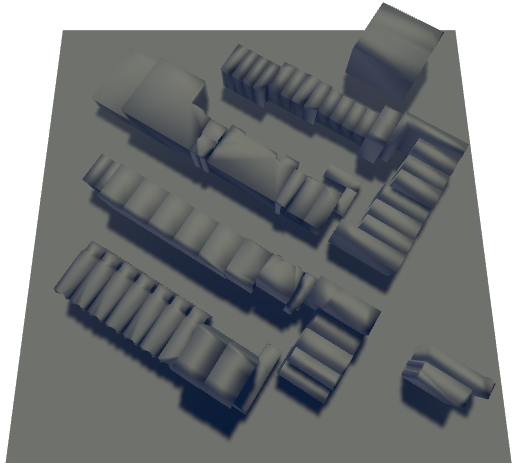}%
		}
		\subfigure[] {\label{c}\centering\includegraphics[width=0.33\columnwidth]{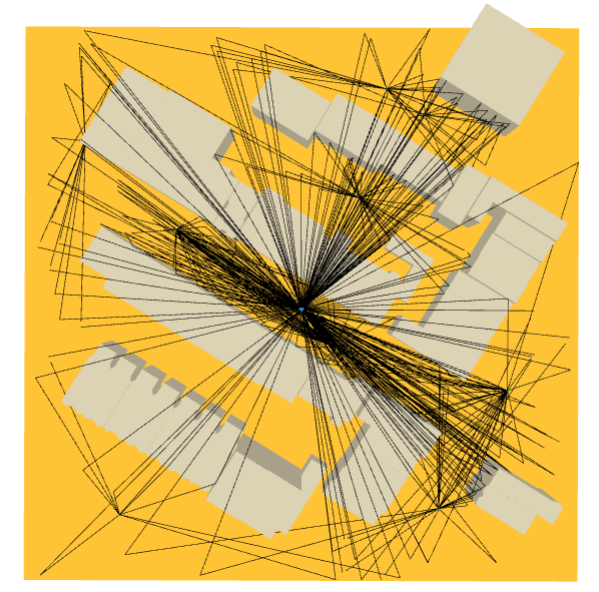}%
		}
		\caption{For a location in New York: (a) street-view image, (b) 3D building distribution, and (c) ray-tracing simulation.}
		\label{fig:NewYork}
	\end{figure}

	To ensure physical consistency, the framework enforces constraints at both the modeling and simulation levels.
	At the receiver side, the region of interest is discretized into a 2D grid, and occluded or indoor points are automatically excluded via geometric visibility detection to prevent non-physical paths. At the propagation level, \textit{Sionna}\footnote{https://github.com/NVlabs/sionna} is used to explicitly resolve the multipath set for each transceiver pair, incorporating line-of-sight (LoS), reflection, diffraction, and scattering mechanisms. Each resolved path contains complex gain, ToA, angular features, and propagation type, supporting cross-scale and cross-scenario physical modeling consistency. 
	Fig. \ref{fig:NewYork} illustrates key intermediate steps of the workflow for a location in New York, including street-view imagery, 3D building distributions, and ray tracing simulations generated via \textit{OpenStreetMap}, \textit{Blender}, and \textit{Sionna}.

	Both transmitters and receivers employ parameterized antenna models (e.g., isotropic radiators or vertically polarized dipoles), ensuring comparability across scenarios while facilitating extensibility for directional array and polarization modeling. This flexible design allows seamless extension to beamforming, RIS-assisted, or MIMO-based systems.

	To further guarantee reproducibility and scalability, OpenPathNet operates under a configuration-driven mechanism where all hyperparameters are managed via unified files. Unlike conventional static datasets, OpenPathNet functions as an open-source dynamic generator, empowering researchers to reproduce experiments or extend simulations to new city maps, frequency bands, and system configurations. Through this holistic framework, OpenPathNet achieves high-fidelity, path-level, and physically interpretable data generation, spanning from city geometry to multipath representation.

	\subsection{Data Generation Pipeline and Implementation}

	\begin{figure}
		\centering
		\includegraphics[width=0.9\columnwidth]{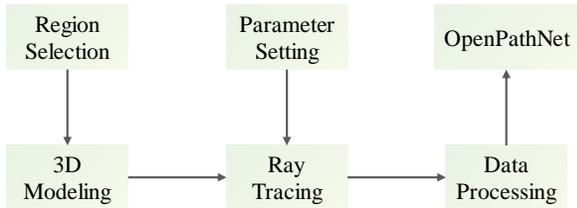}
		\caption{Data generation pipeline of OpenPathNet.}
		\label{fig:pipeline}
	\end{figure}
	The data generation pipeline of OpenPathNet is illustrated in Fig. \ref{fig:pipeline}. A centralized YAML configuration file specifies all essential hyperparameters, including carrier frequency, maximum tracing depth, receiver distribution, transceiver heights, and antenna parameters.

	During the scene construction and geometric preprocessing stage, 3D urban models are generated either by importing real-world scenes based on geographic coordinates or by procedurally synthesizing randomized scenes.
	The target area is then populated with a structured receiver (RX) grid. To ensure geometric precision and computational efficiency, acceleration structures utilizing \textit{Mitsuba} are integrated to detect and exclude points located within building meshes, retaining only visible outdoor receiver points for propagation simulation. This preprocessing step significantly reduces computational overhead while maintaining the physical validity of samples.

	In the core ray tracing stage, the preprocessed scenes and transceiver configurations are loaded into Sionna, which resolves complete multipath sets, covering LoS, reflection, diffraction, and scattering effects.
	For large-scale scenarios or GPU-constrained environments, OpenPathNet introduces an adaptive computation strategy where receiver points are processed in batches. Upon detecting resource bottlenecks, the system dynamically modulates computational density (e.g., adjusting ray sampling count or tracing depth) while upholding a predefined physical fidelity threshold.
	This strategy greatly enhances stability and scalability in large-scale simulations, ensuring consistent outputs across diverse hardware setups. 
	Complementing this algorithmic strategy, OpenPathNet employs a dual-mode architecture that fully leverages parallel processing on high-performance GPUs while enabling flexible distributed computation on CPUs. This design ensures seamless compatibility ranging from individual workstations to large-scale cluster simulations.

	In the post-processing and export stage, ray tracing results for each receiver point are refined. The strongest $N$ paths (e.g., $N=5$) are retained based on channel gain, and key parameters such as ToA, complex gain, and angles are normalized. 
	To cater to diverse research requirements, the framework supports multiple export formats, including CSV and Pickle for detailed analysis, NumPy (.npy) for rapid loading, and auto-generated visualizations for immediate quality inspection.

	Importantly, OpenPathNet is explicitly designed for cross-scenario transferability and algorithmic generalization. Leveraging a unified configuration, researchers can efficiently switch between urban scales, carrier frequencies, and antenna types to systematically evaluate algorithm robustness across diverse propagation environments. In summary, through rigorous configuration management, geometric preprocessing, and robust ray tracing computation, OpenPathNet establishes a high-fidelity, scalable, and reproducible framework, serving as a robust foundation for advancing AI-driven wireless communication research.

	\section{Dataset Characteristics and Analysis}

	This section provides an overview of the core structure, physical attributes, and statistical properties of the accompanying dataset released with OpenPathNet. Unlike conventional simulation-based datasets, it contains explicit path-level multipath parameters generated through realistic geographic modeling and physically consistent constraints. Crucially, the dataset is fully reproducible via the open and configurable OpenPathNet generation pipeline. The resulting large-scale multipath database supports realism, comparability, and scalability by preserving geometric fidelity at the macroscopic level while maintaining statistical consistency with electromagnetic propagation at the microscopic level.

	\subsection{Environmental Information}

	OpenPathNet employs high-fidelity environmental modeling by reconstructing real-world spatial geometries and material characteristics to facilitate accurate ray tracing. Environmental models are derived from actual city layouts, utilizing \textit{Blender} and \textit{OpenStreetMap} to extract building footprints and height profiles. The resulting 3D models incorporate detailed material attributes, such as concrete, glass, and metal, ensuring both geometric and physical realism.

	The accompanying dataset currently spans over ten representative cities, including Beijing, Shanghai, New York, and London, comprising more than 12,000 independent simulation scenes. Each scene corresponds to a $128\ \text{m} \times 128\ \text{m}$ block, sampled across tens of kilometers around city centers to achieve dense tiling across core, transition, and suburban areas. This hierarchical sampling ensures spatial continuity and representativeness on a city scale.

    Following the unified generation protocol, all city subsets share identical configurations, including coordinate systems, geometric resolution, tracing depth, and physical parameters, thereby ensuring semantic comparability. Each scene is stored in \textit{.xml} format, containing complete 3D geometry, material properties, surface reflectivity, and global coordinates. This format allows researchers to reproduce, extend, or annotate scenes for downstream tasks such as channel prediction or semantic labeling.

\begin{figure}[htbp]
		\centering
		\subfigure[] {\label{scenarios1}\centering\includegraphics[width=0.49\columnwidth]{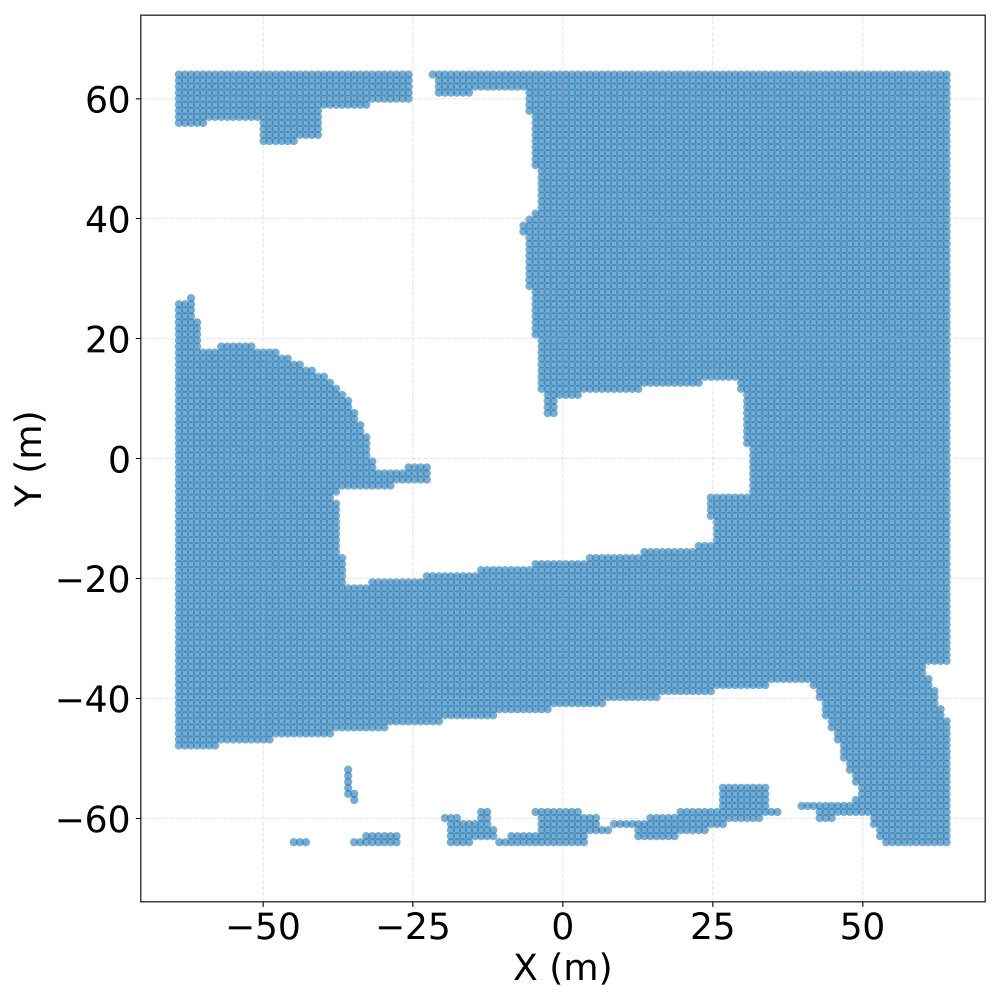}%
		}
		\subfigure[] {\label{scenarios2}\centering\includegraphics[width=0.49\columnwidth]{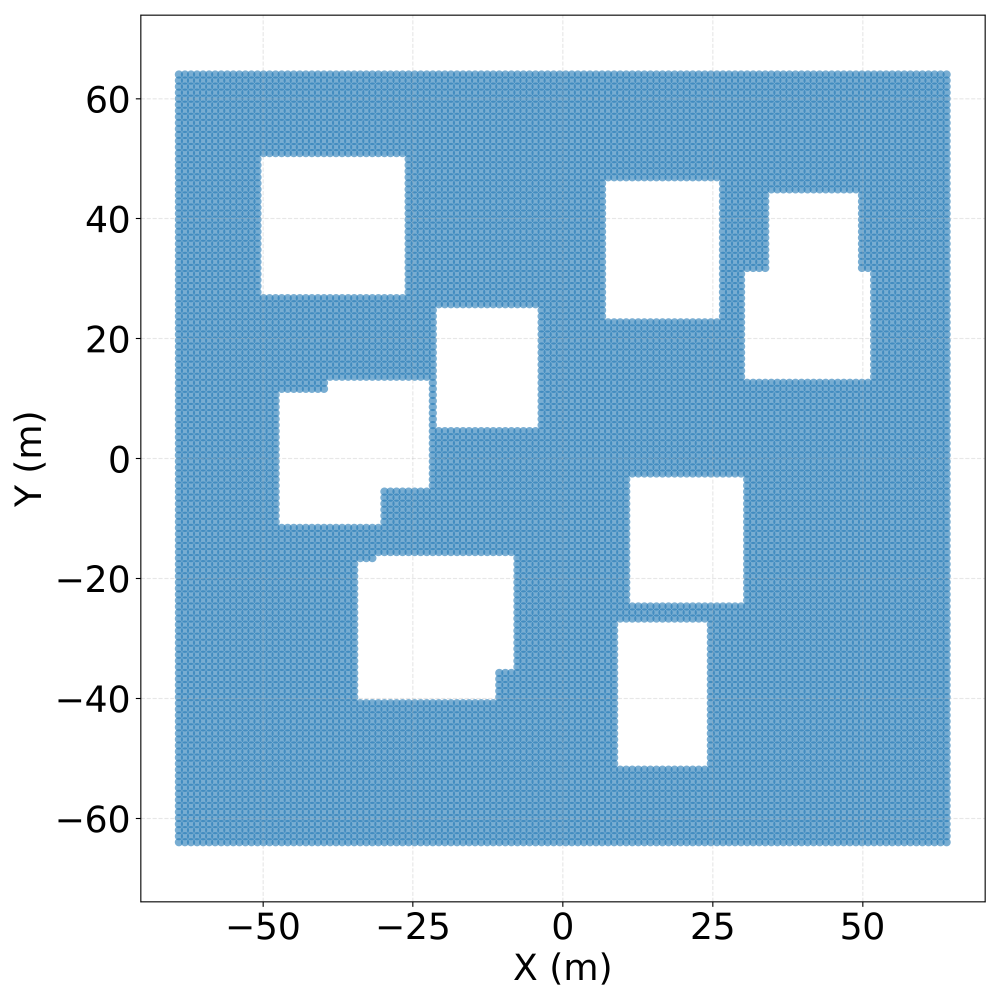}%
		}
		\caption{Examples of building outlines: (a) real-world scenario, (b) randomly generated scenario.}
		\label{fig:scenarios}
	\end{figure}

	In addition, OpenPathNet provides metadata for each scene, recording geographic coordinates, generation parameters, and ray tracing settings to facilitate efficient retrieval and reproducible analysis. Fig. \ref{scenarios1} illustrates a real city environment, while Fig. \ref{scenarios2} depicts a procedurally generated scene for comparison.

	\begin{figure}
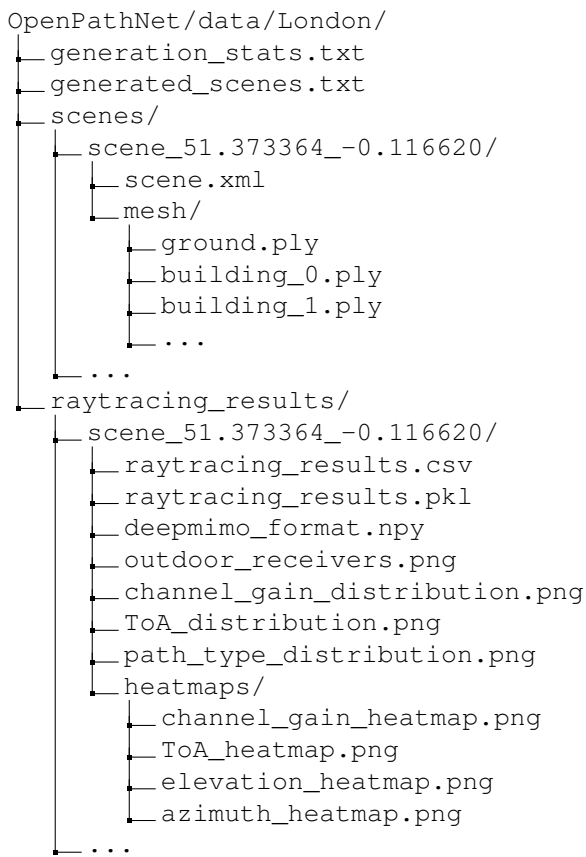

		\dirtree{%
		.1 OpenPathNet/data/London/.
		.2 generation\_stats.txt.
		.2 generated\_scenes.txt.
		.2 scenes/.
		.3 scene\_51.373364\_-0.116620/.
		.4 scene.xml.
		.4 mesh/.
		.5 ground.ply.
		.5 building\_0.ply.
		.5 building\_1.ply.
		.5 ....
		.3 ....
		.2 raytracing\_results/.
		.3 scene\_51.373364\_-0.116620/.
		.4 raytracing\_results.csv.
		.4 raytracing\_results.pkl.
		.4 deepmimo\_format.npy.
		.4 outdoor\_receivers.png.
		.4 channel\_gain\_distribution.png.
		.4 ToA\_distribution.png.
		.4 path\_type\_distribution.png.
		.4 heatmaps/.
		.5 channel\_gain\_heatmap.png.
		.5 ToA\_heatmap.png.
		.5 elevation\_heatmap.png.
		.5 azimuth\_heatmap.png.
		.3 ....
		}
		\caption{Organizational structure of the dataset.}
		\label{fig:directory}
		\vspace{-0.5cm}
	\end{figure}
	
	\subsection{Per-Path Data}

	A key innovation of OpenPathNet is the inclusion of complete per-path channel parameters. Unlike aggregated or channel-level datasets, OpenPathNet preserves detailed multipath structures at each receiver point, providing interpretable and reconstructable channel representations.

	In the current dataset configuration, all scenes use a transmitter at $(0, 0, 30)$ with an omnidirectional antenna. Receivers are uniformly placed on a $128\times128$ grid with $1\text{ m}$ spacing at a height of $1\text{ m}$, enabling dense spatial sampling. The data are stored within a standardized directory structure, as illustrated in Fig. \ref{fig:directory}.

	For each receiver, OpenPathNet provides $N$ path parameters including complex gain, ToA, AoD, angle of arrival (AoA), and path type (LoS, reflection, diffraction, or scattering). Spatial angles are standardized into azimuth $\phi \in [-180^{\circ}, 180^{\circ}]$ and elevation $\theta \in [0^{\circ}, 180^{\circ}]$ to maintain consistency across different environments.

	\begin{figure*}[htbp]
		\centering
		\subfigure[]{\label{fig:heatmap1}\centering\includegraphics[width=0.24\textwidth]{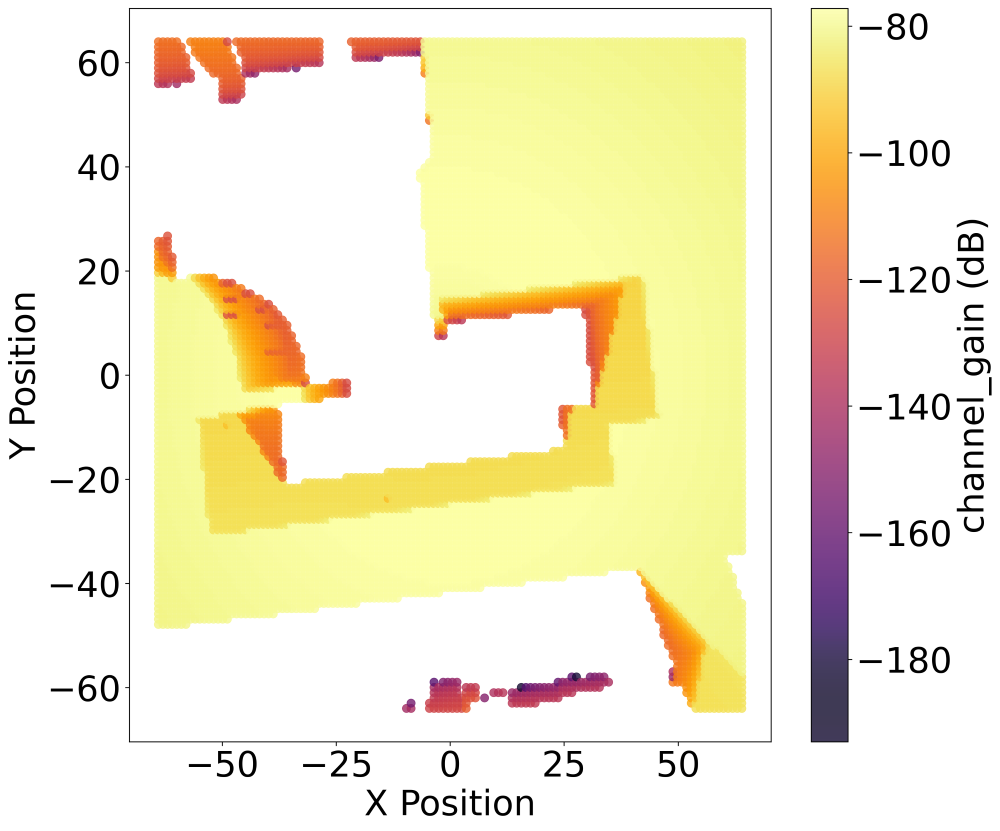}%
		}
		\subfigure[]{\label{fig:heatmap2}\centering\includegraphics[width=0.24\textwidth]{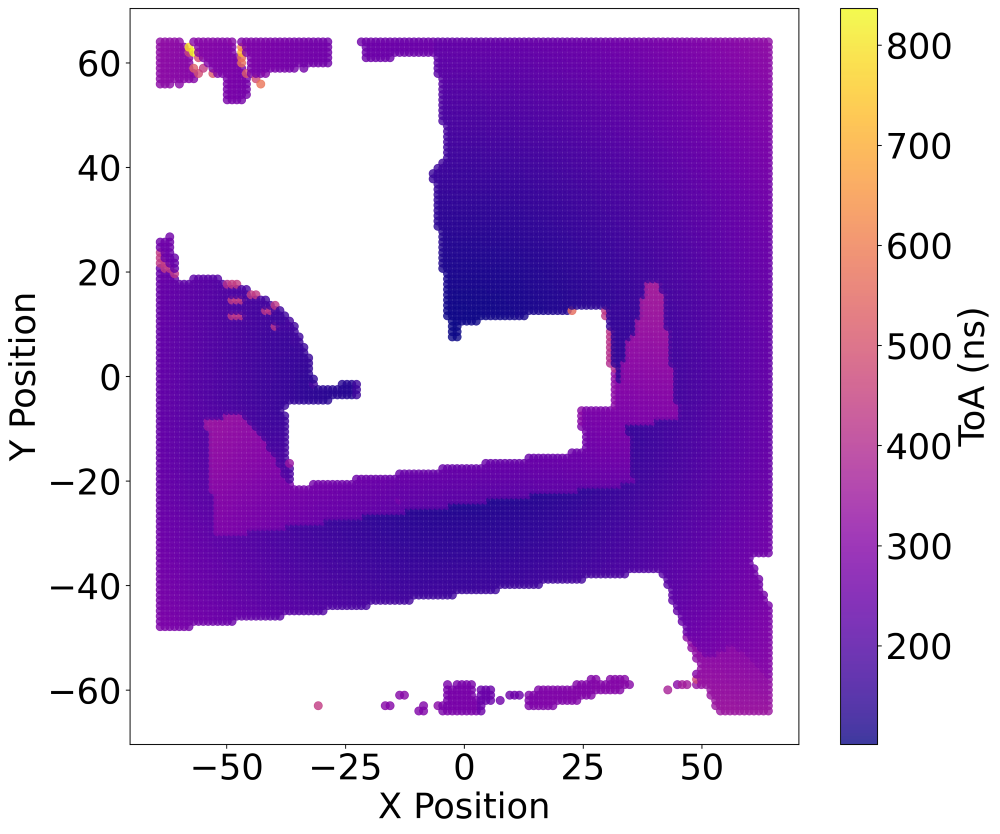}%
		}
		\subfigure[]{\label{fig:heatmap3}\centering\includegraphics[width=0.24\textwidth]{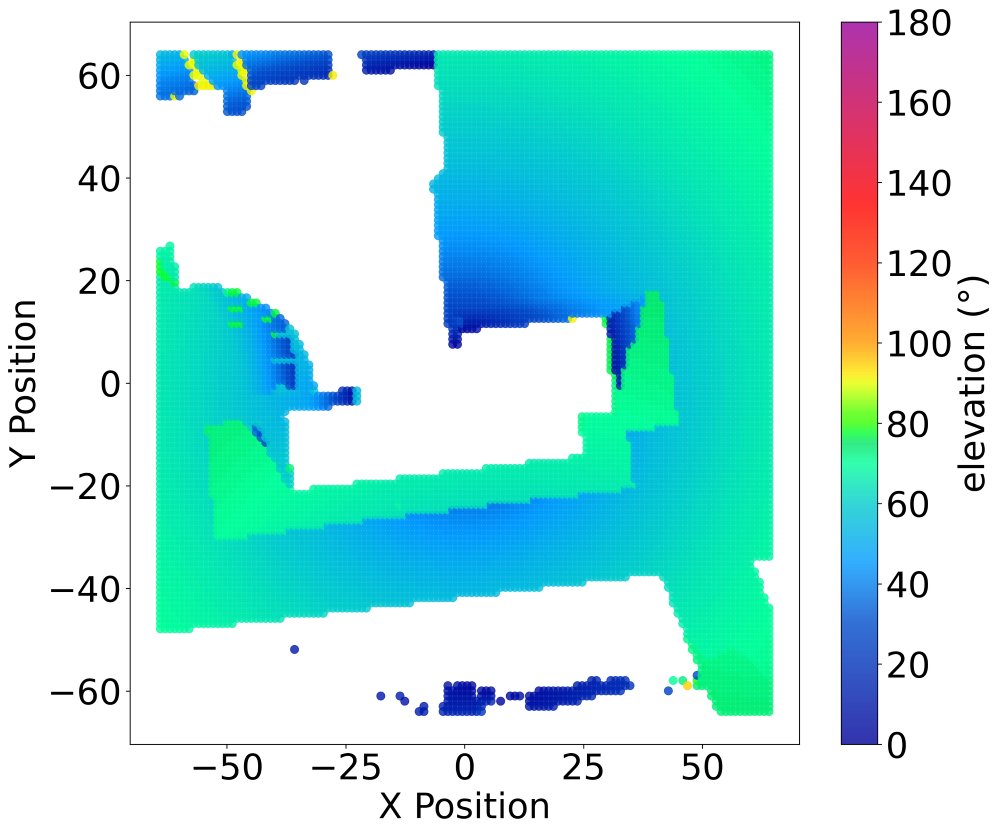}%
		}
		\subfigure[]{\label{fig:heatmap4}\centering\includegraphics[width=0.24\textwidth]{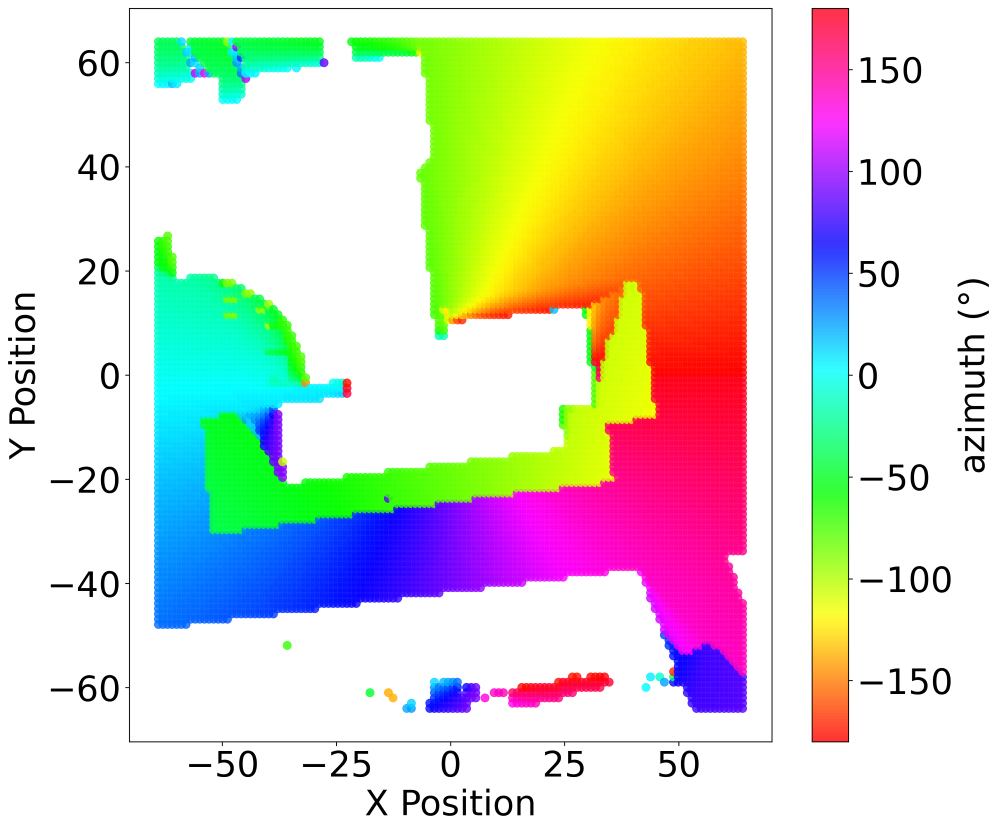}%
		}
		\caption{Visualization of the first-path parameters: (a) channel gain, (b) ToA, (c) elevation angle, and (d) azimuth angle.}
		\label{fig:heatmaps}
	\end{figure*}

	\begin{figure*}[htbp]
		\centering
		\subfigure[]{\label{fig:statistical1}\centering\includegraphics[width=0.34\textwidth]{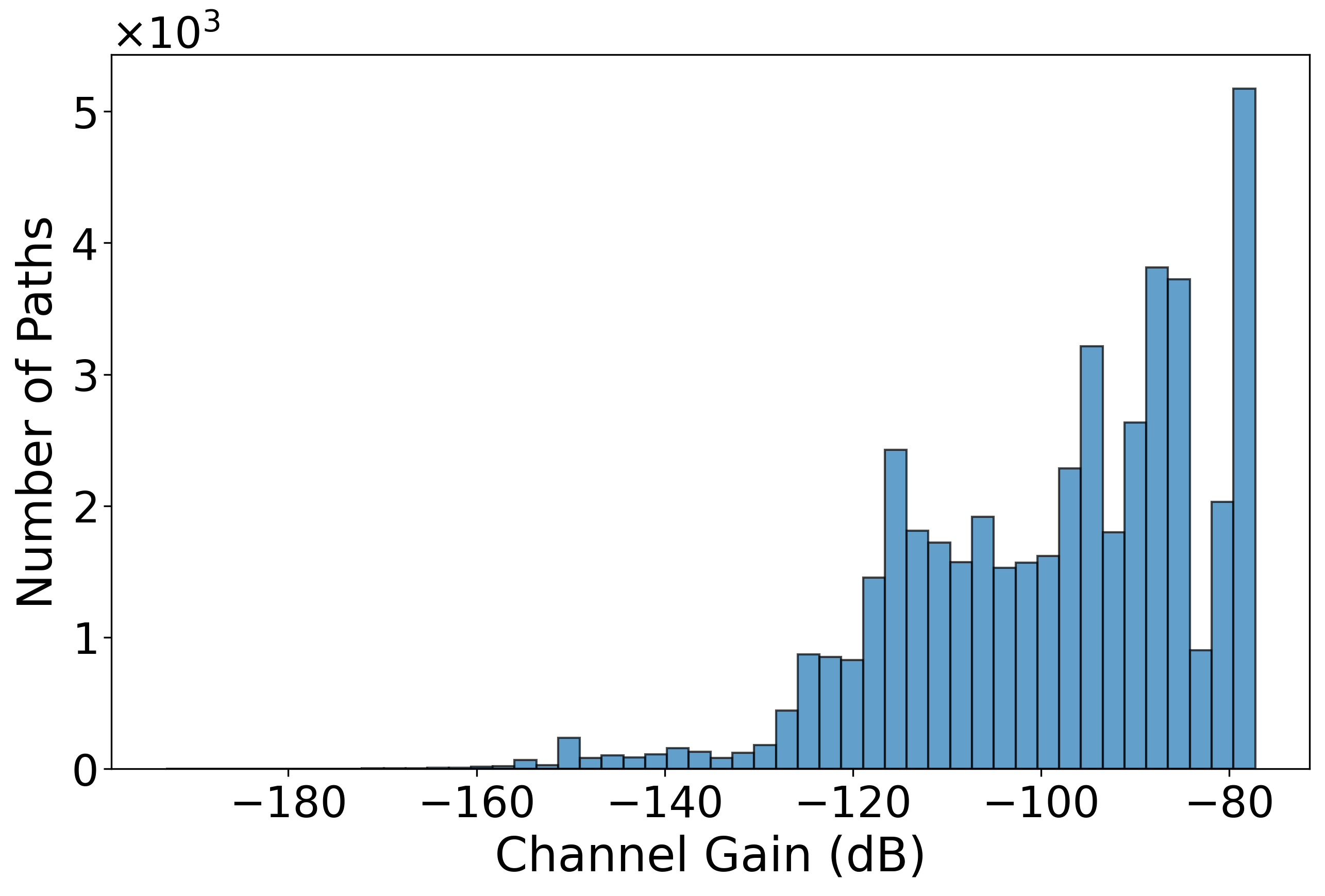}%
		}
		\subfigure[]{\label{fig:statistical2}\centering\includegraphics[width=0.34\textwidth]{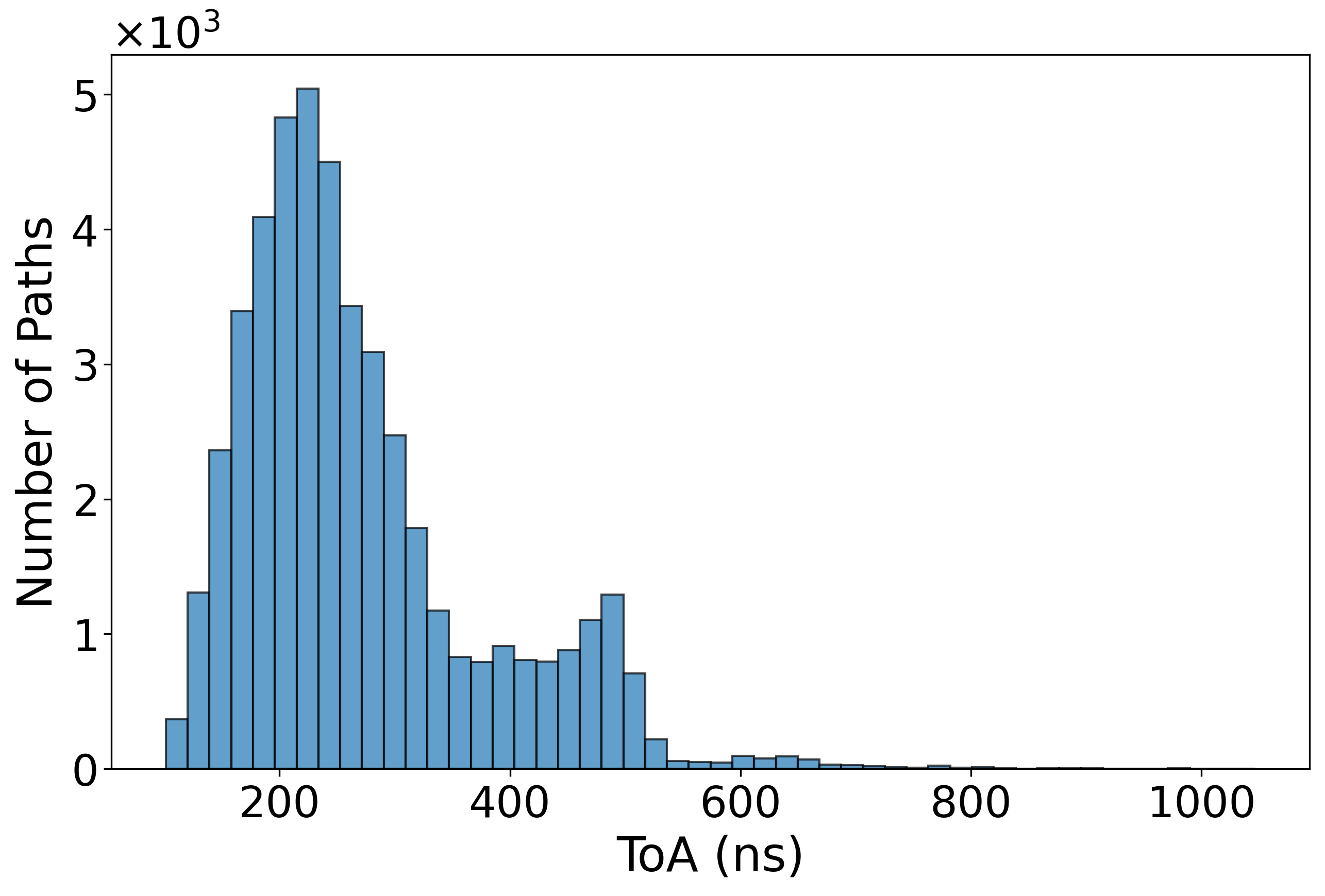}%
		}
		\subfigure[]{\label{fig:statistical3}\centering\includegraphics[width=0.28\textwidth]{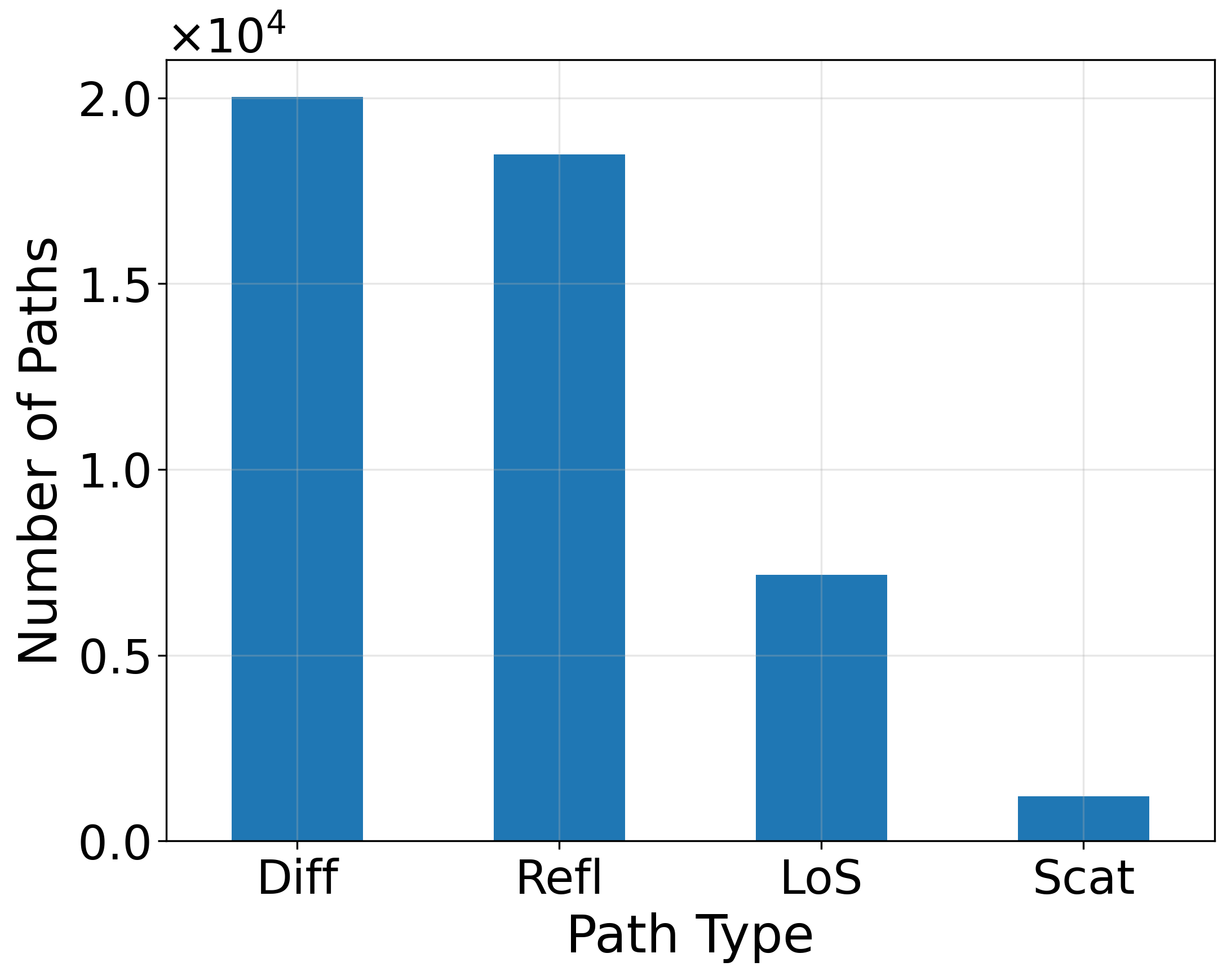}%
		}
		\caption{Example of path-level statistical results in a given scenario: (a) channel gain distribution, (b) ToA distribution, (c) path type distribution.}
		\label{fig:statistical}
	\end{figure*}

	To enhance interpretability, OpenPathNet provides multimodal visualization and statistical analyses. Fig. \ref{fig:heatmaps} presents the spatial distributions of gain, ToA, and the elevation and azimuth angles of arrival for the dominant path. Ideally, power and ToA exhibit radial gradients centered on the transmitter, while angular distributions form ring-like structures consistent with geometric propagation laws. These observations validate the dataset's physical accuracy and demonstrate its capability to capture spatial correlations in electromagnetic propagation, forming a reliable basis for geometry- and vision-based learning tasks. Fig. \ref{fig:statistical} further illustrates the statistical distributions of key physical parameters in the current scenario.

	\subsection{Cross-City Variability and Utility}

	All OpenPathNet city datasets are generated using identical protocols, physical parameters, and quality-control thresholds, including outdoor point filtering and a minimum sample density. This uniform standard ensures that statistical variations across cities are attributable solely to urban morphology rather than implementation or parameter inconsistencies. Consequently, OpenPathNet enables systematic studies on the interplay between city geometry and propagation characteristics.

	Cross-city analyses indicate that the primary differences in multipath statistics arise from building density, street topology, and terrain. Highly dense cities, such as Tokyo and New York, exhibit multimodal distributions in path counts. These variations provide physically interpretable insights into how urban structures shape propagation characteristics.

	OpenPathNet supports two primary research paradigms:  
	(i) \textbf{Intra-City Stratified Analysis}: analyzing path energy and angular statistics across different functional zones, road hierarchies, or density clusters; and  
	(ii) \textbf{Cross-City Generalization Analysis}: comparing metrics such as LoS probability, delay spread, and angular concentration across cities under identical configurations to evaluate AI model generalization and robustness.

	Overall, OpenPathNet combines city-scale geometric constraints with per-path physical fidelity to construct a unified multipath statistical profile. This dual-scale representation captures both global propagation patterns and local variability, providing a credible foundation for multi-domain channel modeling, domain adaptation, and AI generalization benchmarking.

	\section{Conclusion}

	In this paper, we presented OpenPathNet, an open-source, configuration-driven framework designed for scalable and reproducible RF multipath data generation, alongside a comprehensive reference dataset derived from representative real-world environments. OpenPathNet addresses critical limitations in existing RF resources by offering explicit path-level multipath parameters and establishing a transparent, extensible generation pipeline. This design allows researchers to not only reproduce experiments but also systematically expand simulations to new environments and system configurations. Crucially, the framework ensures precise alignment between spatial geometry and propagation modeling, providing geometrically consistent environmental contexts that mirror real-world structures. These explicitly parameterized outputs empower researchers to explore downstream applications, such as data-and-model-driven RF map construction and ISAC. As such, OpenPathNet serves as a robust foundation for reproducible benchmarking in AI-enabled wireless research toward 6G. Future work will extend the framework's coverage to a broader range of cities and frequency bands, while incorporating dynamic environmental factors to capture fine-grained spatiotemporal variations.

\end{document}